# Quantum entanglement between electronic and vibrational degrees of freedom in molecules

Laura K. McKemmish

*School of Chemistry, The University of Sydney, NSW, 2006 Australia*

Ross H. McKenzie

*School of Mathematics and Physics, The University of Queensland, QLD 4072 Australia*

Noel S. Hush

*School of Molecular Biosciences and School of Chemistry, The University of Sydney, NSW, 2006 Australia*

Jeffrey R. Reimers*

*School of Chemistry, The University of Sydney, NSW, 2006 Australia*

\* to whom correspondence should be sent, email reimers@chem.usyd.edu.au

phone +61(2)93514417  fax +61(2)93513329

**Abstract**

We consider the quantum entanglement of the electronic and vibrational degrees of freedom in molecules with a tendency towards double welled potentials using model coupled harmonic diabatic potential-energy surfaces. The von Neumann entropy of the reduced density matrix is used to quantify the electron-vibration entanglement for the lowest two vibronic wavefunctions in such a bipartite system. Significant entanglement is found only in the region in which the ground vibronic state contains a density profile that is bimodal (i.e., contains two separate local minima). However, in this region two distinct types of entanglement are found: (1) entanglement that arises purely from the degeneracy of energy levels in the two potential wells and which is destroyed by slight asymmetry, and (2) entanglement that involves strongly interacting states in each well that is relatively insensitive to asymmetry. These two distinct regions are termed *fragile degeneracy-induced* entanglement and *persistent* entanglement, respectively. Six classic molecular systems describable by two diabatic states are considered: ammonia, benzene, semibullvalene, pyridine excited triplet states, the Creutz-Taube ion, and the radical cation of the "special pair" of chlorophylls involved in photosynthesis. These chemically diverse systems are all treated using the same general formalism and the nature of the entanglement that they embody is elucidated.

I. INTRODUCTION

Entanglement is one of the quintessential 'quantum' phenomena. Here, we develop an understanding of the quantum entanglement between electrons and nuclei in molecules by an analysis of a simple model involving two coupled intersecting potential-energy surfaces. Such a system was first introduced by Horiuti and Polanyi in 1935 to describe proton and hydrogen transfer reactions[1] and was subsequently extended by Hush to oxidation-reduction processes[2] in 1953, now forming the basis of modern electron-transfer theory[3] describing for example exciton and charge transport through molecules, organic conductors and organic photovoltaics as well as electron-transfer reactions in biochemistry.[4] It also describes general racemization processes[5] and has been widely used in spectroscopic analyses, forming the core of Herzberg-Teller theory.[6,7] Throughout *all* of these processes, the vibrational and electronic motions associated with the chemical process become entangled.

In quantum information theory, the simplest type of system is a bipartite pure state[8] given by $|\Psi\rangle = \sum_{ij} c_{ij} |a_i\rangle |b_j\rangle$ where $|a_1\rangle, |a_2\rangle, \ldots |a_n\rangle$ and $|b_1\rangle, |b_2\rangle, \ldots |b_n\rangle$ form any orthonormal set of basis vectors for subsystem A and B respectively. We consider these subsystems as the electronic and vibrational degrees of freedom of a molecule, and the entanglement is, qualitatively, a measure of the connection between them. While many, related, definitions of entanglement have been proposed (see eg.[9-11]), the von Neumann entropy of the reduced density matrix is the most common method of quantifying this entanglement, and we apply it to gain understanding of entanglement in chemical systems.

The chemical model used involves the interaction of a doubly degenerate electronic state with a non-degenerate vibration ($E \otimes B$), a system that can be considered as a special case of Jahn-Teller Hamiltonian[6,7,12] that normally involves the interaction of a doubly degenerate electronic state with a doubly degenerate vibration ($E \otimes E$). It describes molecules with symmetric double-welled potential-energy surfaces such as ammonia and semibullvalene. In these systems, the two wells are equivalent, having the same force constants and energy; these systems form simple models for all processes involving, say, inversion of molecular stereochemistry. We also consider pseudo $E \otimes B$ systems[7] in which an asymmetry $E_0$ is introduced in the relative energies of the two wells, and the general scenario considered is sketched in Fig. 1. In this figure, the red and blue dashed curves represent diabatic potential-energy surfaces describing two non-interacting wells that are coupled by the resonance energy $J$. These give rise to the classic Born-Oppenheimer ground-stated adiabatic potential-energy surface and its associated excited state shown in purple and green in this figure, respectively.

We consider the properties of six molecular examples pertaining to this Hamiltonian: ammonia, semibullvalene, benzene, the Creutz-Taube ion (CT), the bacterial photosynthetic reaction centre radical cation (PRC), and pyridine excited triplet states ($^3$PYR), see Fig. 2 and Table I. These systems display a wide range of qualitative features that control many chemical and physical properties. For ammonia and semibullvalene,[13] $2|J|$ is less than the reorganization energy $\lambda$ required to distort the molecule in one well to the nuclear coordinates of the minimum of other well (see Fig. 1) and so the Born-Oppenheimer surface is double welled.[14] In contrast, for benzene the opposite is true, producing aromaticity with the carbon-carbon bonds taking on equal lengths rather than those expected for a cyclohexatriene. The precise nature of the Creutz-Taube ion[15] has been debated for over 40 years,[16] this molecule being the first mixed-valence compound investigated for which it was apparent that the molecule cannot be simply described as comprising an ion in each of two standard valence states (i.e., an Ru(II) and an Ru(III)), and this molecule became the paradigm through which biological electron transfer processes including those involved in solar-energy conversion during photosynthesis was subsequently interpreted.[17] Molecules showing these types of effects are often classified under the Robin-Day system[18] as either Class II (localized double well), Class III (delocalized single well), or Class II-III (some mixture).[19] Solar to electrical energy conversion occurs in the PRC when the "special pair" of bacteriochlorophylls shown in Fig. 2 ejects an electron to become a dimer radical cation. This ion can be thought of as a "mixed-valence complex" in which each bacteriochlorophyll could take the charge 0 or +1, and, like CT, the charge could alternatively be delocalized over both functionalities. Energy asymmetry in the PRC is induced naturally through asymmetric coordination with the surrounding protein as well as from the asymmetric protein electric field. The excited states of pyridine also display asymmetry but in this case its cause is chemical in origin as nitrogen substitution for CH in benzene makes the associated Kekulé-type structures inequivalent. While this modifies the force constants of the diabatic states as well as their energies, the effect of the energy variation is the most profound and it is realistic to neglect force-constant variations in a simple model description. This scenario is appropriate to a very wide range of excited-state molecular spectroscopy and photochemistry.

We consider only two electronic states coupled by a single vibrational mode, though in general many modes could contribute to the coupling. In practical situations, generalization to multiple modes is typically straightforward and is essential in quantitative analyses. Nevertheless, the essential physics of electron-vibration interaction in molecules can very often be described by the basic two-state one-mode model using appropriately chosen effective parameters. Actually, the one-mode model is a good approximation for most properties of three of the molecules considered herein: ammonia, semibullvalene, and benzene. For the excited triplet states of pyridine,[20] at least 6 modes and 3 electronic states (includes the crossing (n,π*) state) are required in a quantitative analysis, whilst for CT a continuum of solvent modes are critical[16] and 4 electronic states with 70 modes have been used to model PRC.[21] In all cases, some important molecular properties such as the shape, central frequency, and intensity of the characteristic intervalence electronic transition are known to be *independent* of the number of modes used in the analysis.[22] The simple model thus provides a useful general starting position for considering electron-vibration entanglement. The Hamiltonian also describes a superconducting qubit coupled to the resonant microwave mode in a cavity, sometimes called "circuit QED",[23,24] a technology of significant interest for practical quantum information processing.[25]

To manifest this entanglement, we start with an approximate analytical solution for the vibronic wavefunctions of the system obtained using the crude-adiabatic approximation in which the wavefunction is expressed diabatically as a product of an electronic wavefunction, $\phi$, dependent only on electronic coordinates, $r$, and a nuclear wavefunction, $\chi$, dependent only on nuclear coordinates, $R$, i.e.

$$|\psi_{CA}(r,R)> = |\phi(r)>|\chi(R)>. \qquad (1)$$

Exact numerical wavefunctions for the full Hamiltonian are then obtained, using these crude adiabatic wavefunctions as a basis set, in the form

$$|\psi_{exact}(r,R)> = \sum_{ij} c_{ij} |\phi_i(r)>|\chi_j(R)>, \qquad (2)$$

from which the von Neumann entropy of the reduced density matrix can be readily obtained as the electronic and vibrational basis sets are orthonormal.

Such entanglement has been previously quantified[12] in detail for the vibronic ground state of this system for $E_0 = 0$, and some results[26] are also available for $E_0 \neq 0$. Here we examine in detail how the entanglement changes when $E_0 \neq 0$, as well as considering for the first time entanglement within the lowest-energy vibronic excited state. These results bring into prominence the existence of two distinct types of entanglement in the parameter space: entanglement that persists despite the introduction of asymmetry (*persistent entanglement*) and entanglement that disappears (*fragile degeneracy-induced entanglement*). Understanding the consequences of asymmetry thus becomes critical to any application of electron-vibration entanglement to quantum information processing or quantum information transport; particularly influencing the design of experimental apparati and conditions for measuring electron-vibration entanglement. An advanced description of such an apparatus has only been proposed to measure entanglement between the electronic states of pairs of ammonia molecules,[27] but means for measuring related properties for large molecules and nanoparticles with strong environment interactions have been conceived.[28-30] Our results also allow a broad picture to be developed of the significance of entanglement to the understanding of basic chemical bonding and reactivity.

II. MODEL HAMILTONIAN AND ITS PARAMETERS

Expressed in terms of a localized diabatic electronic basis set $|\phi_1>, |\phi_2>$, the couped harmonic diabatic surfaces can be written as a function of a single generalized dimensionless nuclear coordinate $Q$ as

$$\mathbf{H}(Q) = \begin{bmatrix} \frac{\hbar\omega}{2}(Q+\delta)^2 - \frac{\hbar\omega}{2}\frac{d^2}{dQ^2} & J \\ J & \frac{\hbar\omega}{2}(Q-\delta)^2 + E_0 - \frac{\hbar\omega}{2}\frac{d^2}{dQ^2} \end{bmatrix} \quad (3)$$

where $\omega$ is the fundamental vibrational frequency of the two diabatic surfaces (assumed herein to have equivalent force constants), h is Planck's constant, and $\delta$ is the dimensionless displacement of each diabatic surface away from the symmetric configuration. The dimensionless variables $Q$ and $\delta$ can be related to mass-weighted Cartesian-type coordinates using the relationships[31] $Q = q/q_{zpt}$ and $\delta = q_0/q_{zpt}$, where $q$ is the vibrational coordinate in units of $\sqrt{mass} \times length$, $q_0$ indicates the vibrational coordinate where the minima of the diabatic surfaces lie again in units of $\sqrt{mass} \times length$, and $q_{zpt}$ is a scaling factor known as the zero-point vibrational length given by[31] $q_{zpt} = \sqrt{h/\omega}$ in units of $\sqrt{mass} \times length$.

A variety of useful quantities may be described in terms of the basic model parameters. Firstly, the reorganization energy is defined as

$$\lambda = 2\hbar\omega\delta^2 \quad (4)$$

and specifies the energy necessary to distort a molecule in diabatic electronic state 1 to the equilibrium structure of diabatic state 2 and vice versa (see Fig. 1). This energy is often expressed alternatively via the Huang-Rhys factor, $\frac{\lambda}{\hbar\omega} = 2\delta^2$, that indicates the effective number of vibrational quanta excited during electronic transitions between purely localized electronic states. The electronic energy spacing in the presence of the resonance interaction increases to

$$\Delta E = \sqrt{\lambda^2 + 4J^2} \qquad (5)$$

at the geometry of a diabatic minimum for a symmetric ($E_0$=0) system. The Born-Oppenheimer adiabatic approximation provides a good description of the system properties whenever

$$\frac{\hbar\omega}{\Delta E} = 1 \qquad (6)$$

so that the vibrational energy-level spacing is much smaller than the electronic energy-level spacing. This approximation yields the ground-state (GS) and excited-state (ES) electronic potential-energy surfaces shown in Fig. 1 as it allows the effect of the nuclear momentum operator on the electronic wavefunction to be ignored so that **H** can simply be diagonalized parameterically as a function of the nuclear coordinates $Q$. For $E_0 = 0$, the curvatures of these surfaces at $Q = 0$ are given by

$$\frac{\partial^2}{\partial Q^2} E_{GS} = \hbar\omega(1 - \frac{\lambda}{2|J|}) \text{ and } \frac{\partial^2}{\partial Q^2} E_{ES} = \hbar\omega(1 + \frac{\lambda}{2|J|}) \qquad (7)$$

and hence the ground-state surface becomes double welled whenever[14] $\frac{2|J|}{\lambda} < 1$.

The diabatic Hamiltonian Eq. (3) is not unique as all physical properties, including electron-vibration entanglement, are invariant[32] to the rotation of the electronic basis set $|\phi_1\rangle, |\phi_2\rangle$ that produces

$$\mathbf{H'} = \mathbf{RHR}^T \qquad (8)$$

where

$$\mathbf{R} = \begin{bmatrix} \cos\theta & -\sin\theta \\ \sin\theta & \cos\theta \end{bmatrix} . \qquad (9)$$

In particular, a rotation of the localized diabatic Hamiltonian (Eq. 3) typically used to describe double-welled systems such as ammonia by $\theta = 45°$ produces the delocalized diabatic Hamiltonian typically used by spectroscopists to describe the ground and excited states of aromatic molecules.[32]

While four parameters $E_0, \omega, J$ and $\delta$ are specified in Eq. 3, entanglement is independent of the absolute energy scale and hence we simplify the problem by considering only the three independent parameters $2J/\lambda$, $\hbar\omega/\Delta E$, and $E_0/\hbar\omega$. Descriptive parameter values for the iconic systems ammonia, semibullvalene, benzene, CT, PRC, and $^3$PYR are given in Table I.

Figure 3 indicates the physical significance of the ratios $\hbar\omega/\Delta E$ and $2J/\lambda$ by plotting the diabatic and adiabatic potential-energy surfaces generated with $\hbar\omega/\Delta E = 0.1, 1$, or 10 and $2|J|/\lambda = 0.1, 1$, or 10, all at $E_0/\hbar\omega = 0$. As the shapes of the diabatic surfaces are not invariant to electronic-state rotation (Eq. 6), we plot the diabatic states after optimization of $\theta$ to produce the best fit of the lowest 10 vibronic energy levels of the system to the exact energy levels. In this figure, the vibrational levels of the lower-energy diabatic surface are indicated by blue solid lines, while the vibrational levels of the higher-energy diabatic surface indicated by red dashed lines; the optimized angles $\theta$ are also indicated.

An optimized angle near $\theta = 0°$ indicates that fully localized states provide the best-possible diabatic description of the intersecting potential-energy surfaces. This result is produced whenever $h\omega/\Delta E < 1$ and $2|J|/\lambda < 1$, roughly the region in which the ground-state adiabatic potential-energy surface is double-welled and supports below-barrier zero-point vibration. Optimized angles near $\theta = 45°$ result whenever $h\omega/\Delta E < 1$ and $2|J|/\lambda < 1$, the region in which the adiabatic potential-energy surfaces are well-separated from each other and are single welled, typical, say, of the delocalized aromatic states of benzene. In the intermediate region with $h\omega/\Delta E < 1$ and $2|J|/\lambda \sim 1$ the ground-state adiabatic potential energy surface becomes flat and very anharmonic, leading to highly unusual molecular properties, eg., for the Creutz-Taube ion.[16,33] When $h\omega/\Delta E > 1$ the scenario is that of well-separated vibrational levels split by small electronic effects. In this regime the optimum angle is no longer a physically significant indicator, and the effects of vibrations that were not included may dominate any real chemical scenario, bringing into operation say the Jahn-Teller effect[34] that is not included in this simpler one-dimensional approach. As $h\omega/\Delta E$ depicts the ratio of the vibrational and electronic energy-level spacings, the Born-Oppenheimer approximation is expected to perform poorly in this region.

## III. NUMERICAL DETERMINATION OF THE HAMILTONIAN EIGENFUNCTIONS

To find converged numerical solutions to the eigenvectors of **H**, this operator is represented using a product basis of the form $\phi_i \otimes \chi_j$, with $\{\phi_1, \phi_2\}$ forming the (localized) diabatic electronic basis and $\{\chi_1, \chi_2, ..., \chi_n\}$ forming a truncated harmonic-oscillator vibrational basis centered around $Q = 0$. The Hamiltonian matrix elements are then given by:

$$H_{1i,1j} = H_{1j,1i} = \langle \chi_i \phi_1 | \mathbf{H} | \phi_1 \chi_j \rangle = -\delta \hbar \omega \sqrt{\tfrac{i+1}{2}} \delta_{j,i+1} + (i + \tfrac{1}{2})\hbar\omega \delta_{i,j}$$
$$H_{2i,2j} = H_{2j,2i} = \langle \chi_i \phi_2 | \mathbf{H} | \phi_2 \chi_i \rangle = \delta \hbar \omega \sqrt{\tfrac{i+1}{2}} \delta_{j,i+1} + [E_0 + (i + \tfrac{1}{2})\hbar\omega] \delta_{i,j} \quad (10)$$
$$H_{1i,2j} = \langle \chi_i \phi_1 | \mathbf{H} | \phi_2 \chi_j \rangle = J \delta_{i,j}$$

Diagonalizing the Hamiltonian matrix then allows $2n$ independent wavefunctions to be written in the form

$$|\psi\rangle = \sum_{i=1}^{n} c_{1,i} | \phi_1 \chi_i \rangle + c_{2,i} | \phi_2 \chi_i \rangle . \quad (11)$$

In particular, we are concerned with the properties of the ground-state vibronic wavefunction $|\Psi_0\rangle$ and the wavefunction of next highest energy, $|\Psi_1\rangle$.

## IV. VON NEUMANN ENTROPY

The entanglement between the electronic and nuclear degrees of freedom can be expressed as the associated von Neumann entropy,[8,12] obtained by re-expressing the eigenfunctions from Eq. (11) in the form

$$|\psi\rangle = \sum_{k=1}^{2} \phi_k(q) \left[ \sum_{i=1}^{n} c_{k,i} \chi_i(Q) \right] = \sum_{k=1}^{2} \phi_k(q) \chi'_k(Q) \quad \text{or}$$
$$|\psi\rangle = \sum_{i=1}^{n} \left[ \sum_{k=1}^{2} c_{k,i} \phi_k(q) \right] \chi_i(Q) = \sum_{i=1}^{n} \phi'_i(q) \chi_i(Q). \quad (11a)$$

where $q$ and $Q$ are the electronic and vibrational co-ordinates, respectively. Based on these expansions, reduced electronic and vibrational density matrices can then be defined as

$$\rho_{kl}^{E} = \int_{-\infty}^{\infty} \chi'_k(Q)\chi'_l(Q)\,\mathrm{d}Q = \sum_{i=1}^{n} c_{k,i} c_{l,i} \quad \text{and}$$

$$\rho_{ij}^{V} = \int_{-\infty}^{\infty} \phi'_i(q)\phi'_j(q)\,\mathrm{d}q = c_{1,i}c_{1,j} + c_{2,i}c_{2,j}.$$

(12)

While $\rho^E$ is a 2×2 matrix and $\rho^E$ is a $n \times n$ matrix, both matrices share the at-most-two same non-zero eigenvalues $\rho$ and $1-\rho$ (with $0 \leq \rho \leq 1$). The commonality of these two eigenvalues can be seen by writing the quantum state as a Schmidt decomposition, which can have at most two terms,[11] and the von Neumann entropy can be expressed as

$$S = -\rho \log_2 \rho - (1-\rho)\log_2(1-\rho).$$

(13)

If $\rho=0$ or $\rho=1$ then the wavefunction can be expressed as a single product of an electronic wavefunction and a vibrational wavefunction and as a result there is no entanglement, $S=0$. Alternatively, if $\rho=1/2$ then the wavefunction is maximally entangled and $S=1$. It is hence convenient to express the two eigenvalues as

$$\rho = \frac{1}{2}(1-\rho_{\pm}) \quad \text{and} \quad (1-\rho) = \frac{1}{2}(1+\rho_{\pm}).$$

(14)

The entanglement is then maximal when $\rho_{\pm} = 0$ and minimal when $\rho_{\pm} = 1$.

## V. FRAGILE vs. PERSISTENT ENTANGLEMENT

The calculated entanglements depicted over the whole parameter space of the model are shown in Fig. 4 for the ground vibronic-state wavefunction ($S_0$) and first-excited vibronic-state wavefunctions ($S_1$). For symmetric systems ($E_0 = 0$), the entanglement is large whenever $2|J| < \lambda$ and $\Delta E > \hbar\omega$. From Fig. 3 it is clear that this region corresponds to double-well potentials that support strongly localized vibrational motions.

However, Fig. 4 also shows that the introduction of a small amount of asymmetry manifest at, say, $E_0/\hbar\omega = 0.01$ results in a dramatic reduction of the ground-state entanglement, with significant entanglement becoming restricted predominantly to the region with $0.02 < 2|J|/\lambda < 0.5$ and $0.1 < \hbar\omega/\Delta E < 1$. The entanglement in this region is said to be *persistent* whilst that at smaller values of either parameter is said to be *degeneracy-induced* entanglement. Above $E_0/\hbar\omega = 0.1$, the persistent entanglement becomes very weak and maximal in the region near $2|J|/\lambda = 1$, $\hbar\omega/\Delta E < 1$ where the adiabatic ground state is highly anharmonic and sufficiently deep so as to support zero-point motion. Similar results have also been observed using Born-Oppenheimer calculations on a qubit coupled to an oscillator[26] and in applications of the spin-boson model.[35]

The fragility of the degeneracy-induced entanglement can be understood by expanding the non-zero eigenvalues of the ground-vibronic-level density matrix using perturbation theory in the localized limit of $2|J|/\lambda = 1$ (n.b., the resulting equations remain useful even up to at least $2|J|/\lambda = 0.5$):

$$\rho_{\pm} = \left(1 - \frac{1-F_{00}}{1+E_0/(2|J|F_{00})}\right)^{1/2} \qquad (15)$$

where

$$F_{00} = \exp\frac{-\lambda}{2\hbar\omega} \qquad (16)$$

is the Franck-Condon overlap of the two localized-well harmonic-oscillator diabatic ground-state functions. The entanglement is large when $\rho_\pm = 1$, requiring

$$E_0 < 2|J|\exp\frac{\lambda}{2\hbar\omega}. \tag{17}$$

As $\lambda ? 2\hbar\omega$ for any double-welled molecule for which the Born-Oppenheimer approximation is realistic, situations in which $|J|$ is sufficiently small to allow Eq. (17) to hold despite environmentally induced asymmetries are difficult to envisage, except perhaps for molecules in the gas phase at very low temperatures. For example, if very long molecules are employed to take advantage of the usual exponential decay of the coupling with length, then the very size of the molecules means that the large number of intramolecular vibrational modes present will act as a bath, simulating a significant environmental contribution to the asymmetry $E_0$.

A very rapid change in the nature of the eigenvectors of **H** at $2|J|/\lambda$ just $< 1$ occurs and so for $2|J|/\lambda \geq 1$ the extreme sensitivity of the entanglement to $E_0$ is lost and all entanglement is persistent. In this region the eigenvalues of the ground-vibronic-level density matrix can be expanded using perturbation theory based on the delocalized diabatic Hamiltonian that becomes exact in the limit $2|J|/\lambda ? 1$:

$$\rho_\pm = 1 - \frac{\hbar\omega\lambda}{2(\hbar\omega - 2|J|)^2(1 + E_0^2/16J^2)} \tag{18}$$

As the entanglement becomes large when $\rho_\pm = 1$, large entanglement requires both $\hbar\omega \to 2|J|$ and $\hbar\omega/\Delta E = 1$, with in particular for $\hbar\omega/\Delta E = 1$ the entanglement being dominated by the contribution from the lowest density-matrix eigenvalue

$$S \sim -\rho \log_2 \rho, \quad \rho \sim \frac{1}{4}\frac{\lambda}{2|J|}\frac{\hbar\omega}{\Delta E}\left(1+\frac{1}{2}\frac{\lambda}{2|J|}\right). \tag{19}$$

This equation indicates that the entanglement falls off roughly inversely proportionally to $2|J|/\lambda$ when $2|J|/\lambda \gg 1$ and $\hbar\omega/\Delta E = 1$.

Also shown in Fig. 4 are the Born-Oppenheimer potential-energy surfaces corresponding to the parameters that facilitate maximum entanglement for $\Delta E \neq 0$ (for $\Delta E = 0$ the point at $2|J|/\lambda = 0.1$, $\hbar\omega/\Delta E = 0.1$ in the strongly entangled region is illustrated); in addition, the corresponding vibrational density determined from the exact wavefunction is also shown. For $\Delta E / \hbar\omega = 0$ or 0.01, the potential is doubled welled with each well being sufficiently deep to support zero-point vibration, resulting in density profiles that are bimodal (i.e., have a local minimum at or near $Q = 0$ separating two local maxima)[36] as well as associated large entanglements of $S_0 = 0.99$ and 0.93, respectively. For $\Delta E / \hbar\omega \geq 0.1$, the density is unimodal, however, displaying only a single local maximum and no local minima, and the associated entanglement is significantly reduced. On Fig. 4, the regions in which the ground-vibronic-state density is bimodal are indicated and can be seen to correspond to regions in which the entanglement is in excess of ca. 0.7. The qualitative change associated with the unimodal-bimodal changeover has some similarities to *quantum phase transitions* which occur in systems with an infinite number of degrees of freedom.[37] Consideration of the density thus provides a possible means for understanding entanglement within the ground vibronic wavefunction. Similar results occur in other types of qubits, see e.g.[10,23,38]. In particular, systems in which bimodal density profiles survive finite temperature[30,39] are clearly ones in which the entanglement is persistent.

For the first excited-state vibronic level, the overall effect is similar to that for the ground vibronic state but the region of persistent entanglement is much larger and survives beyond $E_0/\hbar\omega = 1$. For the symmetric double-well situation with $E_0/\hbar\omega = 0$ and $2|J|/\lambda < 1$, the first excited vibronic eigenfunction is very similar to the associated ground-state function except for the opposite phasing of the two localized-diabatic ground states, and so the entanglements $S_0$ and $S_1$ behave very similarly in this region, with Eq. 16 again depicting the origin of fragile symmetry-induced entanglement when $\Delta E/\hbar\omega < 0.1$. The similarity between the behavior of $S_0$ and $S_1$ is lost when $E_0/\hbar\omega = 1$, however, as in this scenario the localized diabatic ground-state from one well becomes degenerate with the first excited level of the other, as illustrated in Fig. 5; as a result, $S_1$ remains large at all values of $\Delta E/\hbar\omega$. This resonance also influences $S_1$ for $2|J|/\lambda > 1$, with perturbation theory applied in the limit of $2|J|/\lambda ? 1$ indicating that $S_1$ is large whenever

$$1 \approx \left(\frac{E_0}{\hbar\omega}\right)^2 + \left(\frac{2J}{\hbar\omega}\right)^2. \tag{20}$$

This equation also indicates the effects of an additional resonance that occurs in the delocalized diabatic limit at $2|J|/\hbar\omega = 1$ in which the first excited vibrational level of the ground delocalized state equals the zero-point level of the excited delocalized state; this gives rise to the large persistent entanglement near $\hbar\omega/\Delta E = 1$ for $2|J|/\lambda > 1$, see Fig. 5.

## VI. RELATION TO THE BOHM-AHARONOV TEST OF THE EINSTEIN-PODOLSKY-ROSEN PARADOX

The Einstein-Podolsky-Rosen paradox[40] involves a pair of entangled states of continuous variables (position and momentum). Bohm and Aharonov[41] re-formulated a test of the EPR paradox in terms of discrete degrees of freedom, a pair of singlet-coupled spins. The chemical systems considered are intermediate between EPR and Bohm-Aharonov because one degree of freedom is discrete (the two state electronic system) and the other involves a continuous degree of freedom (the vibrational coordinate). For coupled spins, interactions with the environment destroy entanglement by "measuring" the spin state, whereas in a chemical system interactions with the environment destroy symmetry to prevent entanglement from ever developing; to do this in a Bohm-Aharonov experiment, the environment would need to provide a magnetic fields strong enough to make a component of $^3A_2$ the ground-state of the molecule to eliminate the entanglement. Dissociating a singlet state provides just one example of the general chemical effect of static electron correlation, the type of electron correlation that arises when symmetry rather than interaction determines key features of a quantum state.[42] Static electron correlation can indeed be very fragile, being destroyed by either small intramolecular or external fields that break the symmetry.

## VI. RELATION TO TUNNELLING AND ENERGY TRANSFER KINETICS

Charge and energy may tunnel through symmetric double wells, and ammonia has been considered as a model system for these types of complex phenomena.[43] The effect of minute asymmetry in blocking coherent charge and energy transport has been described.[44] Thus, there appears an intrinsic connection between entanglement and these fundamental chemical processes, and a general theory connecting this effect to the Rabi and Golden-rule limits of kinetics in extended systems has been developed.[45]

**VII. APPLICATIONS TO SOME MODEL MOLECULAR SYSTEMS**

On Fig. 4 the locations in the parameter space appropriate for the 6 paradigm molecular systems described in Table I are indicated, and the calculated values of the ground vibronic level entanglement $S_0$ and the first-excited vibronic level entanglement $S_1$ are indicated in that table. Ammonia, benzene, CT, and semibullvalene all appear on the $E_0/h\omega = 0$ diagram and are found bunched together, despite their qualitatively different chemical properties; this is indicative of the broad range of chemical systems that can be modeled. Benzene, CT, and semibullvalene fall in a region of persistent entanglement, there being no significant change in $S_0$ and $S_1$ at $E_0/h\omega = 0.01$ for these molecules, but the entanglements for ammonia decrease by 70% with this small amount of asymmetry. For ammonia, the value of $h\omega/\Delta E = 0.01$ may be too small for robust operation of some quantum device. At $E_0/h\omega = 0$, ammonia presents the largest ground-state entanglement for any of the molecules considered of $S_0 = 0.43$, however, though the largest such persistent entanglement is 0.37 for CT. PRC is also located in a region of persistent entanglement, but with $E_0/h\omega = 0.6$ its magnitude is restricted to just $S_0 = 0.17$. As expected, the entanglements for the first excited vibronic levels are significantly larger than those for the ground level.

Often it is possible to use chemical or spectroscopic means to modify the basic molecular parameters, opening up the possibility of dynamically switching entanglement on and off. CT not only presents the largest persistent entanglement of all the molecules considered but also allows for this possibility. X-ray photoelectron spectroscopy (XPS) can be used to create a core hole that, because of the small overlap between the core orbitals on the two Ru atoms, becomes 98% localized onto one of the two metal centers,[14] introducing an asymmetry[46] of $E_0$ = 22000 cm$^{-1}$ making $E_0/\hbar\omega$ = 28. The calculated entanglement for this switched CT is $S_1$ = 0.0007, a reduction by a factor of 500.

## VIII. CONCLUSIONS

Double-welled chemical systems by their nature embody electron-vibrational entanglement, and the entanglement of the lowest-energy eigenfunctions is found to be largest when the ground-sate wavefunction has a bimodal density profile, with each maximum depicting localization of the wavefunction on one side of the double well. Entanglement, however, is more widespread than this, forming also in delocalized systems such as the ground state of benzene, though of much reduced magnitude. Nevertheless, entanglement in aromatic molecules can be increased in excited states. Classic dimeric systems such as the Creutz-Table ion and the special-pair radical cation in natural photosynthesis also manifest significant entanglement.

The amount of entanglement manifested in these chemical systems and its sensitivity to environmental effects would appear to limit applicability in quantum information processing and quantum communication. In particular, the fragility of the entanglement for a large range of double-well systems will be critical in limiting performance, though entanglement could possibly be made manifest on a short time scale in a useful way in such systems.[47] Molecules with the greatest likelihood of manifesting persistent ground-state entanglement would need critical parameters of $0.02 < 2|J|/\lambda < 0.8$ and $0.5 < \hbar\omega/\Delta E < 3$. Persistent ground-state entanglement is large when the probability density is bimodal, containing local maxima corresponding to each localized structure. Bimodal behavior requires $2|J|/\lambda < 1$ (to create a double well), $\hbar\omega/\Delta E < 1$ (to trap zero-point vibration within each well), and $E_0 < 2|J|\exp\frac{\lambda}{2\hbar\omega}$ (to allow the localized zero-point levels to interact significantly with each other, requiring at least $E_0/\hbar\omega < 0.5$). Entanglement of the first vibrational level occurs over much the same region of the parameter space as does entanglement of the ground state, developing from the oppositely phased linear combination of the localized-well zero-point vibrations, but also occurs in a much larger region of the parameter space owing to specific new resonances involving locally excited vibrations. While resonance-driven entanglement is persistent, its manifestation would be much more difficult for most systems and also it would be sensitive to the multi-modal nature of real molecular motions.

Ammonia, a molecule previously considered as a candidate for use in quantum information processing,[27] is found to have useful but fragile entanglement and so any application would require extreme isolation from the environment, a challenge that is perhaps inconsistent with the demands of implementing quantum gates, initialization, and readout. Other ammonia-like molecules could have more attractive properties, however, and a thorough scan of candidates is required before any quantum information processor based on electron-vibration entanglement is experimentally investigated. Large molecular systems such as CT and PRC have more attractive parameters than those for the small molecules considered, but these would present great difficulty in experimentally realizing entanglement owing to their large size and strong, sometimes functionally critical, environmental interactions; nevertheless, entanglement can still persist in such environments.[28,30,39] A spectroscopic means by which entanglement could be switched on and off is also described for CT.

In summary, the properties of electron-vibration chemical entanglement are found to be very similar to those for coherent energy transfer, with fragile vibrational resonances being key to both processes. There does exist, however, regions of the parameter space in which the entanglement is persistent, being much less sensitive to external environmental effects, so electron-vibration entanglement has aspects that parallel fragile electron-electron entanglement associated with symmetry (static correlation) and persistent electron correlation associated with (dynamical correlation).

**Acknowledgments**

We thank the National Computational Infrastructure (NCI) for providing computational resources for this project. This work was supported by the Australian Research Council. We thank S. Olsen and X. Huang for helpful discussions.

**TABLE I.** Estimates of parameters values for the coupled harmonic potential-energy surfaces of some different molecular systems, along with the deduced vibronic entanglements $S_0$ and $S_1$.

| System | Ref. | $E_0$ (cm$^{-1}$) | $\hbar\omega$ (cm$^{-1}$) | $J$ (cm$^{-1}$) | $\lambda$ (cm$^{-1}$) | $\dfrac{2|J|}{\lambda}$ | $\dfrac{\hbar\omega}{\Delta E}$ | $\dfrac{E_0}{\hbar\omega}$ | $S_0$ | $S_1$ |
|---|---|---|---|---|---|---|---|---|---|---|
| Ammonia | | 0 | 1700 | -60000 | 160000 | 0.8 | 0.01 | 0 | 0.43 | 0.44 |
| Benzene | | 0 | 1300 | -25000 | 40000 | 1.3 | 0.02 | 0 | 0.07 | 0.16 |
| $^3$PYR | 20 | 2100 | 1620 | -2700 | 16400 | 0.3 | 0.09 | 1.3 | 0.03 | 0.21 |
| CT | 16 | 0 | 800$^a$ | -2900$^b$ | 7100$^b$ | 0.8 | 0.10 | 0 | 0.37 | 0.61 |
| PRC | 21 | 556 | ~980$^c$ | -1020 | 1120 | 1.8 | 0.4 | 0.6 | 0.18 | 0.56 |
| Semibullvalene | 13 | 0 | 1549 | -20800 | 45800 | 0.9 | 0.03 | 0 | 0.22 | 0.37 |

$^a$ This value corresponds to the vibrational frequency of the libration mode of water, which calculations[20] indicate to be the primary carrier of the distortion.

$^b$ These values are debated.

$^c$ This value of the vibrational frequency is a one-mode approximation to the 70 modes used in quantitative simulations.[48]

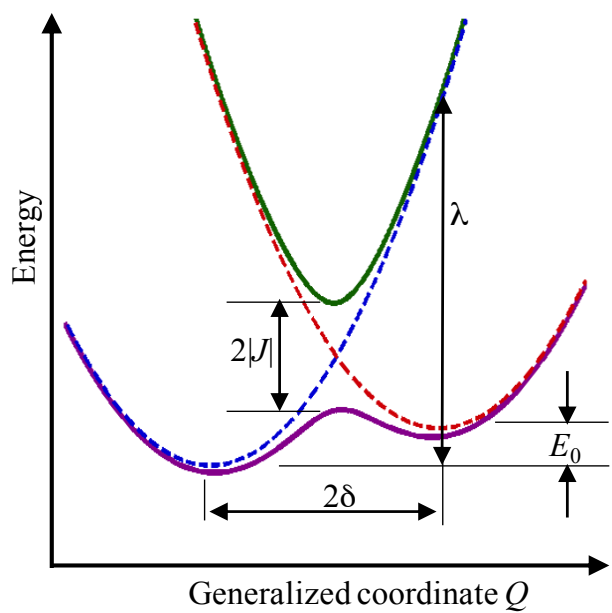

FIG. 1. The chemical model used to describe electron-vibration entanglement. Harmonic diabatic surfaces (blue and red dashed lines), located at minima separated by $2\delta$ in some generalized dimensionless vibrational coordinate with energy asymmetry $E_0$, are coupled by a resonance interaction $J$. The ensuing Born-Oppenheimer ground-state and excited-state adiabatic potential-energy surfaces are denoted by purple and green solid lines, respectively. The reorganization energy $\lambda$ is also indicated.

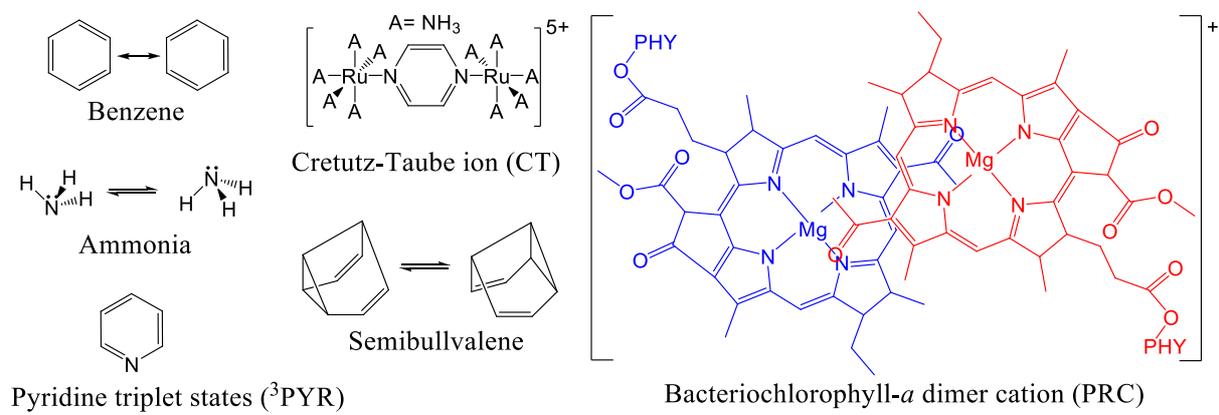

FIG. 2. Some sample molecular systems with electronic states that can be described using two coupled diabatic potential-energy surfaces.

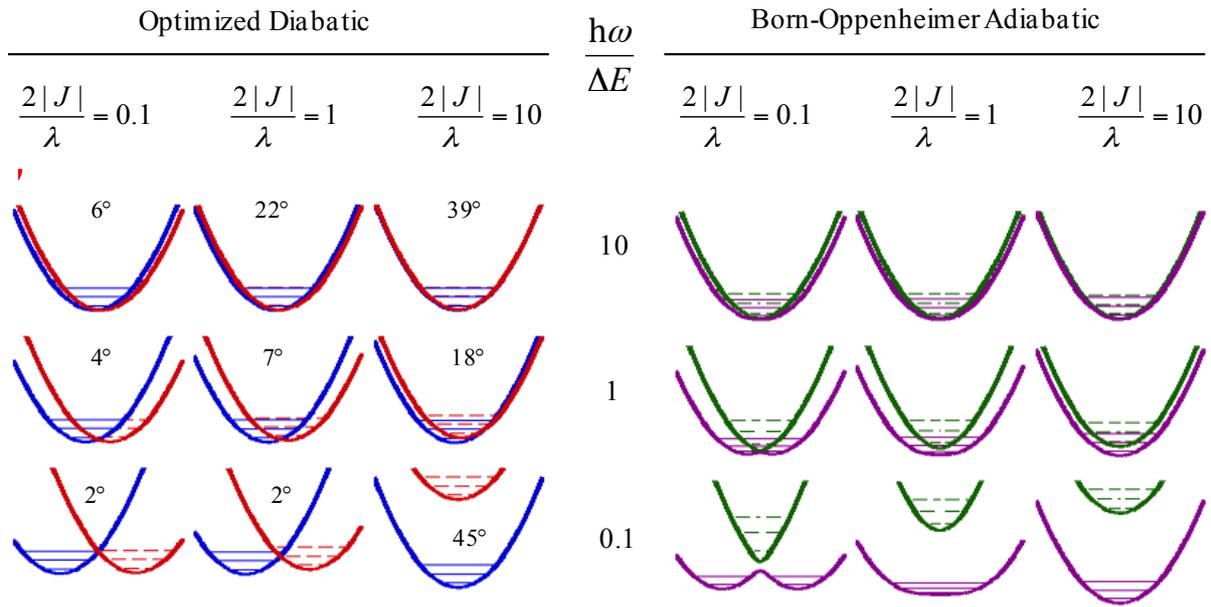

FIG. 3. The optimized diabatic and adiabatic molecular potential energy surfaces (Energy vs. $Q$) and associated lowest energy levels in each state for different values of $2|J|/\lambda$ and $\hbar\omega/\Delta E$; the optimized rotation angles are indicated for the diabatic surfaces.

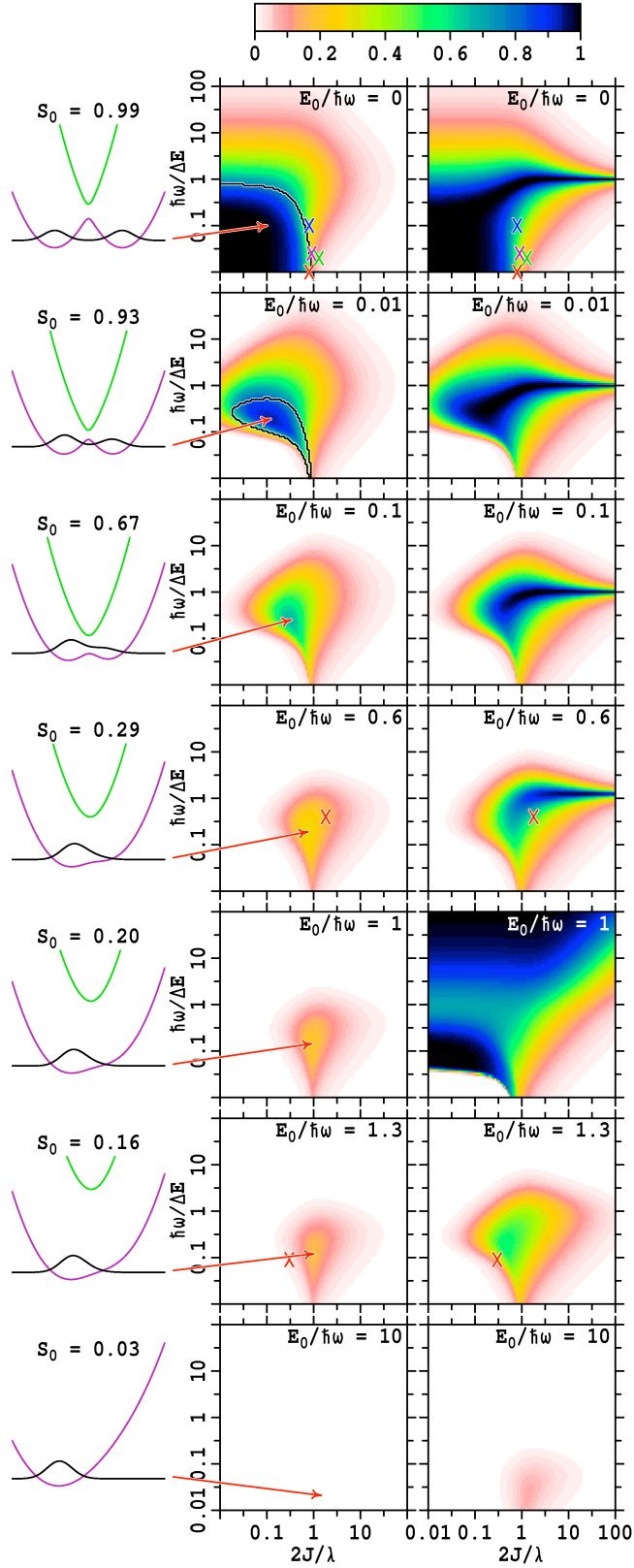

FIG. 4. Persistent versus degeneracy-induced entanglement. Center and right: Electron-nuclear entanglement is shown as a contour plot for the vibronic ground state ($S_0$, centre) and the first vibronic excited state ($S_1$, right) wavefunctions vs. $2|J|/\lambda$ and $\hbar\omega/\Delta E$ at various values of $E_0/\hbar\omega$. The black lines demark regions in which the ground-state vibrational probability density is bimodal or unimodal, while the crosses indicate parameter values relevant to (see Table 1): red- ammonia, $^3$PYR, PRC; green- benzene, blue- CT, purple- semibullvalene. Left: the Born-Oppenheimer potential-energy surfaces (purple and green) and the ground-state vibrational probability density (black) at the indicated parameters.

$$\frac{E_0}{\hbar\omega} = 1 \qquad \frac{\hbar\omega}{\Delta E} = 1$$

FIG. 5. Origin of the resonances that enhance entanglement $S_1$ in the first excited vibronic state: Left- crude adiabatic surfaces at $E_0/\hbar\omega = 1$ and $\hbar\omega/\Delta E = 0.1$ at $\theta = 0$; Right- crude adiabatic surfaces at $\hbar\omega/\Delta E = 1$ and $2|J|/\lambda = 10$ at $\theta = 45°$.